\documentstyle[twoside,fleqn,epsf,espcrc2]{article}

\newcommand{\AmS}{{\protect\the\textfont2
  A\kern-.1667em\lower.5ex\hbox{M}\kern-.125emS}}

\input epsf.sty
\title{Test of the Kugo-Ojima Confinement Criterion\\ in the Lattice Landau Gauge}
\author{Hideo Nakajima\address{Department of Information Science, Utsunomiya University,\\
2753 Ishii, Utsunomiya 321-8585 Japan (e-mail nakajima@is.utsunomiya-u.ac.jp)} and   
        Sadataka Furui\address{School of Science and Engineering, Teikyo University, \\
1-1 Toyosatodai, Utsunomiya 320-8551, Japan (e-mail furui@dream.ics.teikyo-u.ac.jp)}}
\begin{document}
\begin{abstract}
We present the first results of numerical test of the Kugo-Ojima confinement
criterion in the lattice Landau gauge.  The Kugo-Ojima criterion of colour
confinement in the BRS formulation of the continuum gauge theory is
given by $u^a_b(0)=-\delta^a_b$, where 
\[\int dx e^{ip(x-y)}\langle 0|TD_\mu c^a(x)g(A_\nu\times \bar c)^b(y)|0\rangle
=(g_{\mu \nu}-{p_\mu p_\nu \over p^2})u^a_b(p^2).\quad(*)\]
We measured the lattice version of
$u^a_b(0)$ in use of $1/(-\partial D(A))$ where $D_\mu(A)$ is a lattice
covariant derivative in the new definition of the gauge field as $U=e^A$.
We obtained that $u^a_b(0)$ is consistent with $-c\delta^a_b,\ c=0.7$ in
$SU(3)$ quenched simulation data of $\beta =5.5$, on $8^4$ and $12^4$.
We report the $\beta$ dependence and finite-size effect of $c$.
\end{abstract}

\maketitle

\section{INTRODUCTION}

The colour confinement problem in the continuum gauge theory
was extensively analysed in use of the
BRS formulation by Kugo and Ojima\cite{KO}. 

The QCD lagrangian is invariant under the BRS transformation and the physical space is specified as 
the one that satisfies the condition
 ${\cal V}_{phys}=\{|phys\rangle\}$ 
\[
Q_B |phys\rangle=0.
\]
where
\[
Q_B=\int d^3x\Big[
B^aD_0c^a-\partial_0B^a\cdot c^a+\displaystyle{i\over2}g
\partial_0 \bar c^a\cdot(c\times c)^a
\Big]
\]
and $(F\times G)^a=f_{abc}F^b G^c$.  

Under the assumption that {\bf BRS singlets have positive metric},
it is proved that ${\bf {\cal V}_{phys}}$ has positive semidefinite
in such a way that {\bf BRS quartet
particles appear only in zero norm}.

One finds from the BRS transformation
that for each colour $a$, a set of massless asymptotic fields
$\chi^a, \beta^a, \gamma^a, \bar \gamma$ form a 
BRS quartet.

 The Noether current corresponding to the conservation of the colour 
symmetry is $gJ^a_\mu={\partial ^\nu}{F^a_{\mu\nu}}+\{Q_B, D_\mu \bar c\}$,
where its ambiguity by divergence of antisymmetric tensor should be understood,
and this ambiguity is utilised so that massless contribution may be eliminated
for the charge, $Q^a$, to be well defined.

 Denoting $g(A_{\mu}\times \bar c)^a\to u^a_b\partial_{\mu}\bar \gamma^b$,
and then $D_{\mu}\bar c^a\to (1+u)^a_b\partial_{\mu}\bar \gamma^b$,
one obtains the eq.(*)
provided $A_{\mu}$ has a vanishing expectation value.
The current $\{ Q_B,D_\mu \bar c\}$ contains the massless component,
$(1+u)^a_b\partial_\mu \beta^b(x)$.
We can modify the Noether current for colour charge $Q^a$ such that
\[
gJ'^a_\mu=gJ_\mu-{\partial ^\nu}{F^a_{\mu\nu}}=\{Q_B, D_\mu \bar c\}.
\]
In the case of ${\bf 1+u=0}$, massless component in $gJ'_0$ is vanishing and
the colour charge 
\begin{equation}
Q^a=\int d^3x \{Q_B, g^{-1}D_0\bar c^a(x)\}
\label{kg}
\end{equation}
becomes {\bf well defined}.

The physical state condition  $Q_B |phys\rangle=0$ together with the 
equation (\ref{kg}) implies that all BRS singlet one particle states
 $|f\rangle \in {\cal V}_{phys}$ are colour singlet states.
This statement implies that all coloured particles in ${\cal V}_{phys}$
belong to BRS quartet and have zero norm. This is the {\bf colour
confinement}.

\section{LATTICE CALCULATION OF $u^a_b$}

The Faddeev-Popov operator is
\begin{equation}
{\cal M}[U]=-(\partial\cdot D(A))=-(D(A)\cdot\partial),
\end{equation}
where the new definition of the gauge field is adopted as $U=e^A$,
and the lattice covariant derivative $D_{\mu}(A)=\partial_{\mu}+Ad(A_{\mu})$
is given in \cite{NF}.

 The inverse, ${\cal M}^{-1}[U]=(M_0-M_1[U])^{-1}$, is calculated
 perturbatively by using the Green function of the  Poisson equation 
$M_0^{-1}=(-\partial^2)^{-1}$  and $M_1=\partial_\mu Ad (A_\mu(x))$, as
\begin{equation}
{\cal M}^{-1}=M_0^{-1}+\sum_{k=0}^{N_{end}}(M_0^{-1}M_1)^kM_0^{-1}.
\end{equation}

 In use of colour
source $|\lambda^a x\rangle$ normalised as
$Tr \langle \lambda^a x|\lambda^b x_0\rangle=\delta^{ab}\delta_{x,x_0}$,
the ghost propagator is given 
by
\begin{equation}
G^{ab}(x,y)=\langle Tr \langle \lambda^a x|({\cal M}[U])^{-1}|
\lambda^b y\rangle \rangle
\end{equation}
where the outmost $\langle\rangle$
specifies average over samples $U$. 

The ghost propagators of $\beta=5$ and $5.5$ are almost the same and they are
infrared divergent which can be parameterised as $\displaystyle {1\over p^{2.2}}$.
We observed that the ghost propagators of $\beta=6$ is similar to that of $\beta=5.5$ and its
finite-size effect is small\cite{SS}. 
 
In the similar way, one can calculate the Kugo-Ojima parameter
at $p=0$ as,
\begin{eqnarray}
(g_{\mu \nu}-{p_\mu p_\nu \over p^2})u^a_b(p^2)|_{p=0}\qquad\qquad\qquad\qquad\qquad\nonumber\\
=\langle Tr \langle \lambda^a p|D_{\mu}(A)({\cal M}[U])^{-1}(Ad(A_{\nu}))|
\lambda^b p\rangle \rangle|_{p=0}
\end{eqnarray}

We observed that off-diagonal element of $u^a_b$ is consistent to zero, 
but there are statistical fluctuations. 
 The projection operator $g_{\mu\nu}-{p_\mu p_\nu\over p^2}$ in 
equation (*) 
is treated such that it has  an expectation value ${3\over 4}$ in the limit of 
$p_\mu\to 0$.  

Making the accuracy of the covariant Laplacian equation solver higher, 
we observe the tendency that the expectation value of $|u^a_a|$ increases.

\begin{table*}[htb]
\setlength{\tabcolsep}{.2pc}
\caption{Kugo-Ojima parameter $u^a_a$. Space-diagonal($\mu=\nu$) and off-diagonal components. 
All data of $8^4$ are the average of 100 samples. 'Z3' and 'min' means Z3 twisting and the minimum Landau
gauge fixing.}

\begin{tabular*}{\textwidth}{@{}l@{\extracolsep{\fill}}rrrrrr}

\hline

         & $diag$  & $off-diag$ &  $diag_1$ &$diag_2$ & $diag_3$ & $diag_4$ \\
\hline
$\beta=5.5,8^3\times 16$ &-0.739(135) &   0.002(60) &  -0.776(109) &  -0.779(105) & -0.818(118) &  -0.581(49)\\
$\beta=5.5,12^4$          &-0.715(46) &  0.003(32) & -0.729(60)  &  -0.713(43) & -0.705(39) & -0.712(38)\\
$\beta=5.5,8^4$          &-0.664(69) &  0.002(45) &  -0.669(71) &  -0.656(70) & -0.667(67) & -0.664(67)\\
$\beta=6.0,12^4$          &-0.548(133) &  -0.015(85) &  -0.555(123) &  -0.561(107) & -0.508(133) & -0.566(159)\\
$\beta=6.0,8^4,with\ Z_3$  &-0.303(80) &  0.002(29) &  -0.286(76) &  -0.307(66) & -0.325(81) & -0.293(91)\\
$\beta=6.0,8^4,with\ Z_3, min$  &-0.308(88) & -0.000(35) &  -0.312(123) &  -0.311(78) & -0.317(75) & -0.292(59)\\ 
$\beta=6.0,8^4,no\ Z_3$   &-0.354(176) &  -0.001(76) &  -0.339(130) &  -0.347(161) & -0.378(239) & -0.353(151)\\
$\beta=8.0,8^4,with\ Z_3$  &-0.183(74) & 0.002(20) &  -0.177(71) &  -0.197(77) & -0.221(83) & -0.138(19)\\ 
$\beta=8.0,8^4,no\ Z_3$    &-0.338(513) &  0.0116(251) &  -0.264(278) & -0.359(553) &  -0.334(610) &  -0.394(536)\\

\hline
\end{tabular*}
\end{table*}

\begin{figure}[htb]
\begin{center}
\epsfysize=120pt\epsfbox{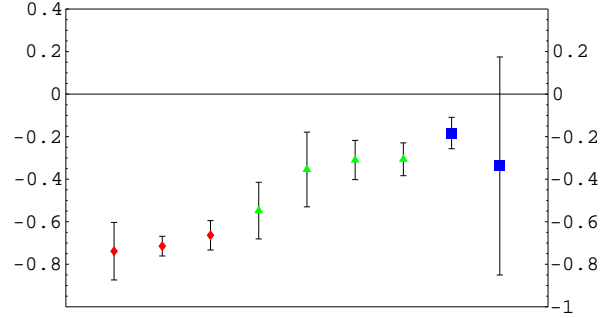}
\caption{The dependence of space and colour diagonal part of the Kugo-Ojima parameter $u^a_a$ on
$\beta$ and lattice size. (An average over the four directions and eight adjoint representations.)
The data points are  $\beta=5.5, 8^3\times 16; \beta=5.5, 12^4;
\beta=5.5, 8^4; \beta=6, 12^4; \beta=6, 8^4(no\ Z_3); \beta=6, 8^4(with\ Z_3); \beta=6, 8^4(with\ Z_3,
minimum\quad Landau); \beta=8, 8^4(with\ Z_3); 
\beta=8, 8^4(no\ Z_3)$ respectively from left to right.}
\label{kugo1}
\end{center}
\end{figure}

At $\beta=8$, direct measurement of $u^a_b$ 
gives a large fluctuation, but suitable $Z_3$ twisting treatment
for each sample
so that the Polyakov scatter plot should be concentrated around
$\arg z=0$, suppresses the fluctuation and makes the quality of the data better.
We consider that this treatment is indispensable in the simulation where
$Z_3$ symmetry persists and the $Z_3$ factor affects the observed quantity.
The similar behaviour is observed in $\beta=6, 8^4$ lattice.
The minimum Landau gauge fixing via smeared gauge fixing performed at $\beta=6,
8^4$ lattice does not change the expectation value obtained after the $Z3$
twisting but reduces the standard deviation.

The absolute value of $u_a^a$ is plotted as the function of the spatial extent of the lattice $a L$ where $a$ is
calculated by assuming $\Lambda_{\overline {MS}}=100MeV$. We find for $a L < 2fm$, there exists large
 finite-size effectDWe expect that by making $L$ large and $a$ small, such that $a L > 2 fm$, the absolute value of 
$u_a^a$ becomes closer to 1.

\begin{figure}[htb]
\begin{center}
\epsfysize=120pt\epsfbox{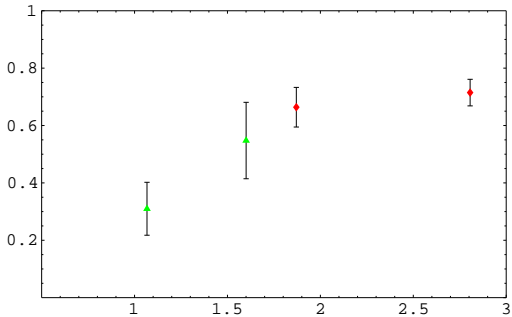}
\caption{The finite-size effect of the  Kugo-Ojima parameter $|u^a_a|$ as the function of the spatial extent 
of the lattice $a L (fm)$.}
\end{center}
\hfil
\end{figure}
 Non-symmetric lattice $8^3\times 16$ yields non-symmetric
data in $\mu$ of (*). 
This fact shows necessity of tuning lattice constants
according to the non-symmetric
lattice size and the lattice dynamics. 

\section{SUMMARY AND DISCUSSION}

Proof of Kugo-Ojima colour confinement is accomplished successfully only in case
of $u^a_b= -\delta^a_b$, and this condition is suggested to be a
necessary condition as well.
We did the first numerical tests of this
criterion by the nonperturbative dynamics of lattice Landau gauge.
We observed that the value at $\beta=5.5$ is around $-0.7$. Its absolute value
decreases as $\beta$ increases.

We observed the gluon propagator is infrared finite\cite{NF} and the ghost 
propagator is infrared divergent, suggested to be more singular than
$\displaystyle{1\over p^2}$, but less singular than $\displaystyle{1\over p^4}$.
These results qualitatively agree with the Gribov-Zwanziger's 
conjecture\cite{Gv,Zw}, and are consistent with the results of Dyson-Schwinger equation
\cite{SHA}.
It is nice to observe that the infrared finiteness of the gluon propagator
is in accordance with the Kugo-Ojima
colour confinement. As stated in their inverse Higgs mechanism theorem,
if we have no massless vector poles in all channels of the gauge field,
$A^a_{\mu}$, and if the colour symmetry is not broken at all,
it follows that $1+u=0$.\cite{Iz}.

This work is supported by High Energy Accelerator Research Organization, 
KEK Supercomputer Project(Project No.99-46), and by Japan Society for the Promotion of Science, Grant-in-aid for Scientific Research(C) (No.11640251).
  
\end{document}